\documentclass[jcp,aip,
 amsmath,amssymb,
reprint,%
]{revtex4-1}

\usepackage{graphicx}
\usepackage{dcolumn}
\usepackage{bm}
\usepackage{color}

\graphicspath{{./figures/}}

\newcommand{\REV}[1]{{\color{black} #1}}

\begin{document}


\title{Self-organization in suspensions of
end-functionalized semiflexible polymers under shear flow} 



\author{Jin Suk Myung}
\email{j.myung@fz-juelich.de}
\affiliation{Theoretical Soft Matter and Biophysics, Institute of Complex Systems and Institute for Advanced Simulation, Forschungszentrum J{\"{u}}lich, 52425 J{\"{u}}lich, Germany}

\author{Roland G. Winkler}
\email{r.winkler@fz-juelich.de}
\affiliation{Theoretical Soft Matter and Biophysics, Institute of Complex Systems and Institute for Advanced Simulation, Forschungszentrum J{\"{u}}lich, 52425 J{\"{u}}lich, Germany}

\author{Gerhard Gompper}
\email{g.gompper@fz-juelich.de}
\affiliation{Theoretical Soft Matter and Biophysics, Institute of Complex Systems and Institute for Advanced Simulation, Forschungszentrum J{\"{u}}lich, 52425 J{\"{u}}lich, Germany}



\begin{abstract}
The nonequilibrium dynamical behavior and structure formation of end-functionalized semiflexible
polymer suspensions under flow are investigated by mesoscale hydrodynamic simulations.
The hybrid simulation approach combines the multiparticle collision dynamics method for
the fluid, which accounts for hydrodynamic interactions, with molecular dynamics simulations for the semiflexible polymers.
In equilibrium, various kinds of scaffold-like network structures are observed, depending on polymer
flexibility and end-attraction strength.
We investigate the flow behavior of the polymer networks under shear and analyze their
nonequilibrium structural and rheological properties.
The scaffold structure breaks up and densified aggregates are formed at low shear rates, while
the structural integrity is completely lost at high shear rates.
We provide a detailed analysis of the shear-rate-dependent flow-induced structures.
The studies provide a deeper understanding of the formation and deformation of network structures
in complex materials.
\end{abstract}

\pacs{}

\maketitle 


\section{Introduction} \label{sec1}

Smart and responsive complex materials can be achieved by self-organization of simple building blocks.  By now, a broad range of  functionalized colloidal and polymeric building blocks have been proposed and designed. \cite{Vermant2005,Glotzer2007,Sciortino2008,Solomon2010,Chen2011,Capone2012,Capone2013,Yi2013} This comprises synthetic  colloidal structures, e.g., patchy or Janus colloids \cite{Chaudhary2012,Yi2013,Walther2013} or biological molecules such as DNA duplexes. \cite{Nakata2007} \REV{These building blocks are able to self-organized into gel-like structures, e.g., hydrogels, which are able to undergo reversible changes in response to external stimuli.\cite{Ilman1991,Tanaka1982,Petka1998,Guo2005,Das2006,Richter2008,Grodzinski2010,Gruhn2013,White2013}
Thereby, rodlike molecules, such as viruses \cite{Reddy2012} or telechelic associative polymers, \cite{Annable1993,Serero1998,Khalatur1999,Berret2001,Pellens2004b,Hosono2007} exhibit novel scaffold-like structures, and theoretical and experimental studies have been undertaken to unravel their structural and dynamical properties in suspensions.} Here, polymer flexibility and end-interactions are the essential parameters to control the properties of the self-assembled network structures. \cite{Chelakkot2006,Myung2014,Taslimi2014}

The appearing structures can be directed and controlled by external parameters, specifically by the application of external fields such as a shear flow.\cite{Vermant2005} Here, a fundamental understanding of the nonequilibrium response of a network structure is necessary for the rational design of new functional materials and that of already existing synthetic and biological scaffold-like patterns. \cite{Claessens2006,Ramos2007,Chelakkot2009,Lieleg2010,Lieleg2011,Broedersz2014}

Computer simulations are an extremely valuable tool to elucidate the self-organized structures of functionalized polymers.  Monte Carlo \cite{Chelakkot2006} and molecular dynamics simulation \cite{Guo2005,Khalatur1999,Myung2014} studies of coarse-grained models of end-functionalized flexible, semiflexible, and rodlike polymers in solution have shown that in thermal
equilibrium self-organized scaffold-like network structures form above a critical attraction strength and within a range of concentrations. This network formation is strongly affected by the polymer flexibility, because flexible polymers can span a larger range of distances between connections points, even form loops, and deform easily thereby generating softer networks. The molecular dynamics simulation studies of telechelic polymers of Ref.~\onlinecite{Khalatur1999} predict flower-like micellar aggregates for flexible polymers. For stiffer polymers, significant morphological changes appear, with liquid-crystalline-like order of adjacent polymers and inter-connected structures.\cite{Khalatur1999,Myung2014} Recent nonequilibrium simulations of end-functionalized rodlike polymers exhibit further structural changes under shear flow. \cite{Taslimi2014} At low shear rates,  the scaffold structure compactifies, while at intermediate shear rates novel bundle-like structures appear with nematically ordered rods. In the limit of very strong flows, all structures are dissolved and the rodlike polymers align in a nematic fashion.

In this article, we extend the previous studies and investigate the influence of shear flow on the scaffold-like network structure of end-functionalized {\em semiflexible} polymers.
Both, the structure properties under shear flow as well as the rheological properties are analyzed for various shear rates. We find that an initial scaffold structure breaks up and densified aggregates are formed at low shear rates, while the structural integrity is completely lost at high shear rates. Thereby, flexibility gives rise to particular compact aggregates at intermediate shear rates.
In addition, the relaxation behavior of shear-induced structures after cessation of flow is analyzed in part in order to elucidate the reversibility of the shear-induced structures.

We apply a hybrid simulation approach, which combines the multiparticle collision dynamics (MPC) method for the
fluid, \cite{Malevanets1999,Ihle2001,Kapral2008,Gompper2009} which accounts for hydrodynamic interactions,
\cite{Tuezel2006,Kapral2008,Gompper2009,Huang2012} with molecular dynamics simulations for the semiflexible
polymers. \cite{Winkler2004,Kapral2008,Huang2010,Huang2013} The MPC method has successfully been applied to study the equilibrium and nonequilibrium dynamical properties of complex systems such as polymers, \cite{Malevanets2000.1,Ripoll2004,Mussawisade2005,Kapral2008,Frank2008,Gompper2009,Huang2010,Huang2013,Jiang2013}
colloids, \cite{Lee2004,Hecht2005,Padding2006,Ripoll2008,Wysocki2009,Whitmer2010,Franosch2011,Singh2011}
vesicles and blood cells, \cite{Noguchi2005,McWhirter2009} as well as various active systems. \cite{Buyl2013,Elgeti2015,Hu2015}

The combination of coarse-grained modeling of end-functionalized polymers and a particle-based mesoscale hydrodynamic simulation technique is ideally suited for such a  study.
On the one hand, we want to elucidate the general principles of structure formation under nonequilibrium conditions. The achieved insight will be useful to understand the behavior of a broad spectrum of experimental systems, ranging from highly flexible synthetic polymers, e.g., telechelics, to stiff biological macromolecules, such as DNA segments. On the other hand, mesoscale hydrodynamic simulation
approaches are essential, because only they allow to reach the large length and time scales, which are required to capture
the long structural relaxation times in shear flow with typical shear rates of $10^1 - 10^3$ Hz. \cite{Vermant2005,Pipe2009}
In addition and most importantly, particle-based hydrodynamic simulation approaches naturally include thermal fluctuations,
which are indispensable for a proper description of polymer entropy and entropic elasticity. Of course, coarse-grained
modeling has its limitations in predicting the behavior of particular experimental systems quantitatively. Here, additional
simulations of atomistic models are required to predict binding energies and bending rigidities.

This paper is organized as follows. The simulation approaches are introduced in Section \ref{sec2}. The deformation of the
polymer network under shear and rheological properties are discussed in Section \ref{sec3}, and the dependence on the polymer
flexibility is addressed. Relaxation of shear-induced structures is discussed as well.
Section \ref{sec4} summarizes our findings.

\section{Model and Simulation Algorithm} \label{sec2}

\subsection{Mesoscale Hydrodynamic Solvent: Multiparticle Collision Dynamics}

Our hybrid simulation approach combines the multiparticle collision dynamics method for the fluid with molecular dynamics simulations for the semiflexible polymers. \cite{Huang2010} In the MPC method, the fluid is represented by ${N_s}$ point particles of mass ${m}$, which interact with each other  by a stochastic process. \cite{Malevanets2000,Kapral2008,Gompper2009} The dynamics proceeds in  two steps---streaming and collision. In the streaming step, the particles move ballistically and their positions are updated according to
\begin{equation} \label{eq:eq1}
  {{\bf{r}}_i}(t + h) = {{\bf{r}}_i}(t) + h{{\bf{v}}_i}(t).
\end{equation}
Here, ${{\bf{r}}_i}$ and ${{\bf{v}}_i}$ are the position and velocity vector of the \textit{i}th particle, and ${h}$ is the time between collisions.
In the collision step, the particles are sorted into cells of a cubic lattice with lattice constant ${a}$, and their velocities are rotated
relatively to the center-of-mass velocity ${{\bf{v}}_{cm}}$ of the cell
\begin{equation} \label{eq:eq2}
  {{\bf{v}}_i}(t + h) = {{\bf{v}}_{cm}}(t) + {\bf{R}}(\alpha )[{{\bf{v}}_i}(t) - {{\bf{v}}_{cm}}(t)],
\end{equation}
where  ${{\bf{R}}(\alpha )}$ is the rotation matrix for the rotation around a randomly oriented axis by the fixed angle ${\alpha}$. The orientation of the axis is chosen independently for every collision cell and collision step.

\subsection{Polymer Model}

A semiflexible polymer is modeled as a linear sequence of  ${N_m}$ mass points of mass  ${M}$. These monomers are connected by harmonic
springs with bond potential
\begin{equation} \label{eq:eq3}
  {U_b} = \frac{k_b }{2}\sum\limits_{k = 1}^{{N_m} - 1} {{{\left( {\left| {{{\bf{r}}_{k + 1}} - {{\bf{r}}_k}} \right| - l} \right)}^2}},
\end{equation}
where ${\bf r}_k$ is the position of monomer $k$, ${l}$ is the equilibrium bond length, and ${k_b}$ is
the spring constant.  Semiflexibility is implemented by the bending potential
\begin{equation} \label{eq:eq4}
  {U_B} = \frac{{{\kappa}}}{2}\,\sum\limits_{k = 2}^{{N_m} - 1} {{{\left( {{{\bf{r}}_{k + 1}} - 2{{\bf{r}}_k} + {{\bf{r}}_{k - 1}}} \right)}^2}}.
\end{equation}
Here, ${\kappa} = k_BT L_p/l^3$ is the bending rigidity, where ${k_B}$ is the Boltzmann constant, ${T}$ is the temperature, and ${L_p}$ is the persistence length.
Excluded-volume interactions between monomers are taken into account by the shifted and truncated Lennard-Jones potential (LJ)
\begin{equation} \label{eq:eq5}
  {U_{LJ}} = \left\{ \begin{array}{ll}
4\varepsilon \left[ {{{\left( \displaystyle{\frac{\sigma }{r}} \right)}^{12}} - {{\left( \displaystyle{\frac{\sigma }{r}} \right)}^6} + A} \right] & r < {r_c} \\
0 & r \ge {r_c} \end{array} \right. ,
\end{equation}
where ${\sigma}$ is the diameter of a monomer and ${\varepsilon}$ is the interaction strength. Aside from the polymer ends, all monomer interactions are purely repulsive, with the cutoff distance $r_c= 2^{1/6}\sigma$ and the shift ${A} = 1/4$. For the attractive ends, the cutoff is set to  $r_c = 2.5\sigma$ and $\varepsilon$ is varied according to the desired attraction strength.
\subsection{Polymer-Solvent Coupling}

The polymer-solvent coupling is implemented by including the monomers in the collision step. Hence, the particle center-of-mass velocity of a cell containing monomers is
\begin{equation} \label{eq:eq6}
  {\bf v}_{cm}(t) = \frac{\sum\limits_{i = 1}^{N_s^c} m{\bf v}_i(t)
   + \sum\limits_{k = 1}^{N_m^c} M{\bf v}_k(t)}{mN_s^c + MN_m^c},
\end{equation}
where ${N_s^c}$ and ${N_m^c}$ are the number of solvent and monomer particles in the cell, respectively. \cite{Gompper2009}

\begin{figure*}
\centering
  \includegraphics[width=\textwidth]{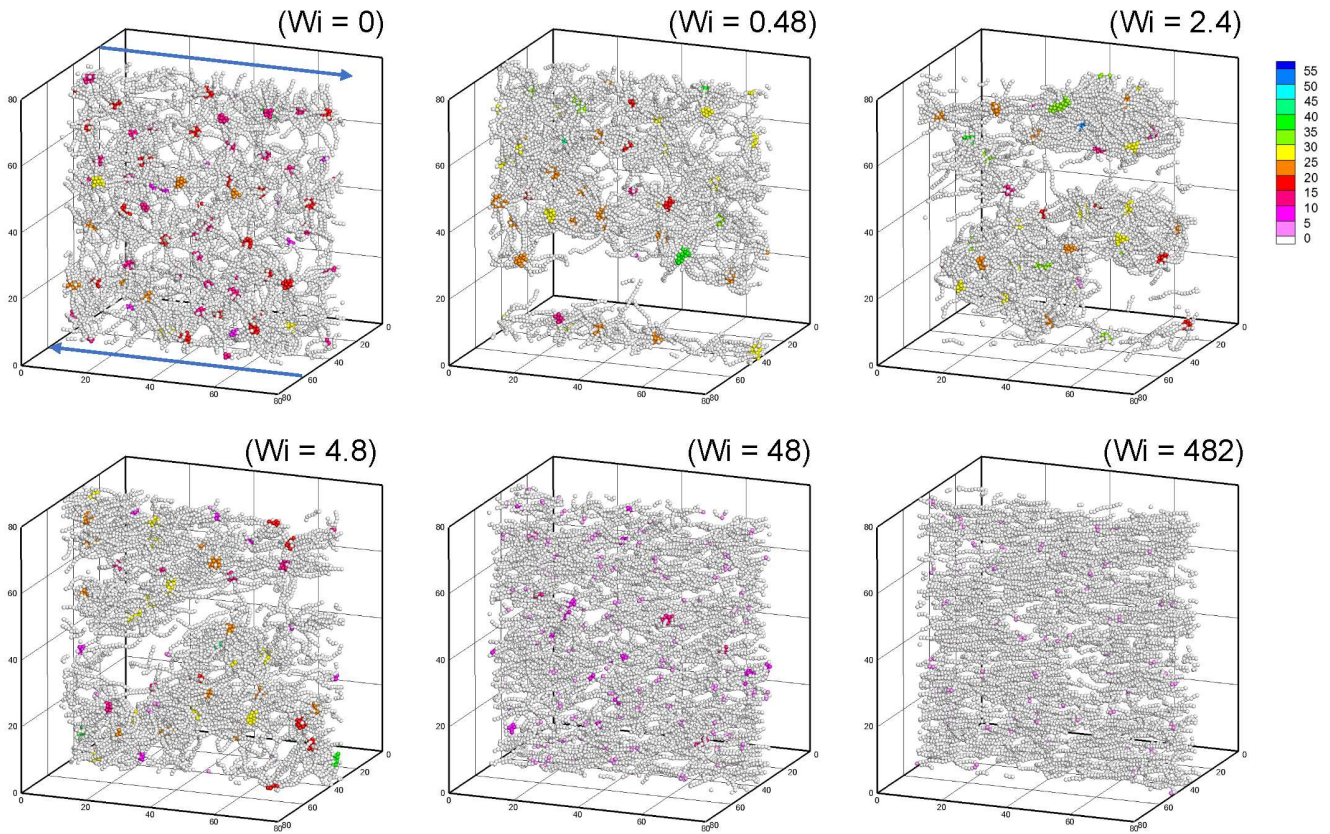}
  \caption{Polymer configurations for the persistence length $L_p/L=1/2$, the
  end-attraction strength $\varepsilon/k_BT=4 $, and for various shear rates.
  Only beads with the slice $30 \le z/a \le 50$ are shown, and the color code corresponds to the number of adjacent ends. The shear direction is indicated by arrows ($Wi=0$). The configurations correspond to the stationary state with the total strain $\gamma = 200$.}
  \label{fgr:shear}
\end{figure*}

\subsection{Simulation Set-up and Parameters}

We consider a cubic simulation box of side length $L_s=80a$. The parameters for the MPC fluid are ${\alpha  = 130^\circ}$, $h = 0.1 \sqrt{ ma^2 /k_BT}$, and the mean number of fluid particles in a collision cell ${\left\langle {N_s^c} \right\rangle  = 10}$. We choose the bond length $l$ as length unit and set for the collision-cell size $a=l$. Moreover, we set $\sigma =l$, $M=10m$, and $k_b =  5000 k_BT/a^2$. The latter ensures that the bond lengths remain close to the equilibrium value even under shear flow for all considered shear rates.
The equations of motion for the monomers are solved by the velocity-Verlet algorithm with  time step $h_p = h/50$. \cite{Allen1987}

In total, 2000 polymers of length $N_m=20$ are considered. Approximating a polymer by a cylinder of length  $L \approx N_m \sigma = 20\sigma$, the polymer-volume fraction is $\phi = 0.06$. Initially, the polymers are distributed randomly in the simulation box and are equilibrated without end-attraction. Then, the end-attraction is turned on and the system is again equilibrated until expectation values reach a steady state.

Shear flow is imposed on equilibrium structures by Lees-Edwards boundary conditions, \cite{Allen1987} with the flow direction along the $x$ axis and the gradient along the $y$ axis of the Cartesian reference system. Shear is characterized by the Weissenberg number $Wi = {\dot{\gamma}\tau}$, where ${\dot{\gamma}}$ is the shear rate and ${\tau}$ is the end-to-end vector relaxation time of a polymer in dilute solution. \cite{Huang2010} Explicitly, the  values of the relaxation time are $\tau/\sqrt{ma^2 /k_BT} = 4815$ and $10220$ for the persistence lengths $L_p/L =1/2$ and $5$, respectively.\cite{Myung2014} \REV{In the following, we will refer to polymers with $L_p/L=1/2$ and $5$ as {\em semiflexible} and {\em rodlike}, respectively.}

For an efficient simulation of the polymer and MPC fluid dynamics, we exploit a graphics-processing-unit (GPU) based version of the simulation code. \cite{west:14}

\begin{figure}
\centering
  \includegraphics[width=8.5cm]{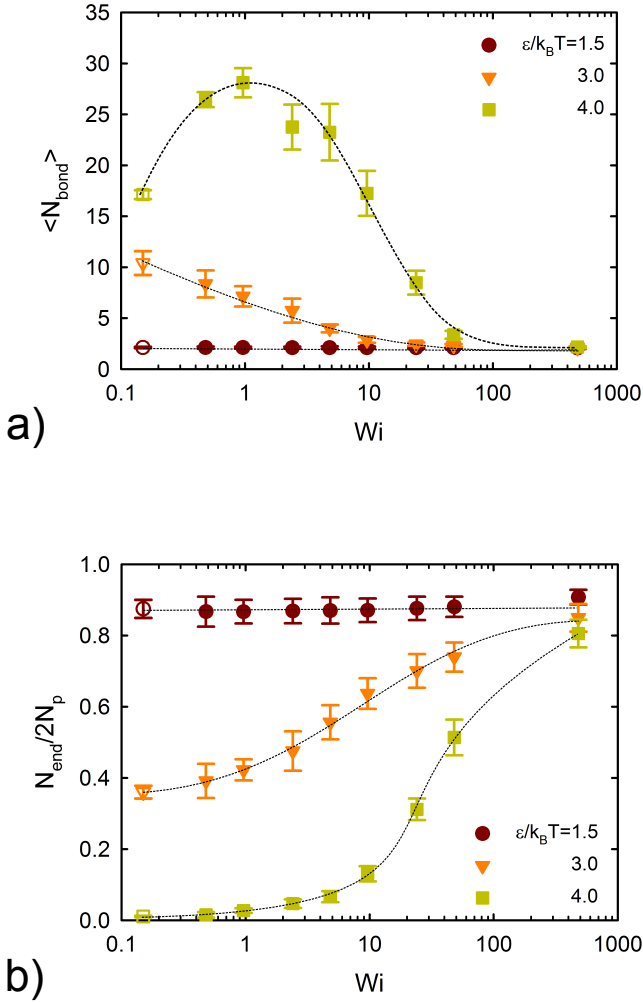}
  \caption{(a) Average coordination number, and (b) number of free
  end-beads as a function of shear rate for the persistence length $L_p/L=1/2$ and
  the end-attraction strengths ${\varepsilon/k_BT=1.5, \ 3, \ 4}$. Open symbols correspond to the numbers at equilibrium without flow.
  Error bars display the magnitude of the fluctuation in the steady state. The ines are guides for the eye.}
  \label{fgr:bond}
\end{figure}

\begin{figure}
\centering
  \includegraphics[width=8.5cm]{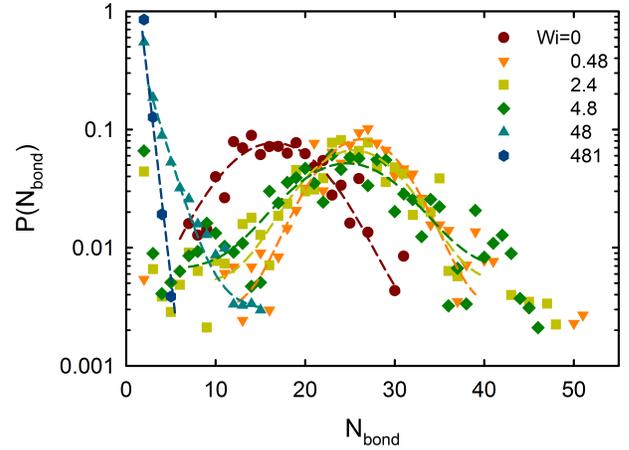}
  \caption{Distributions of the coordination number for the persistence length $L_p/L=1/2$,
  the end-attraction strength $\varepsilon/k_BT=4$ and various shear rates. The dashed lines are fits to guide the eye, with
  a Gaussian function for $Wi \lesssim 5$ and
   an exponential function for $Wi \gtrsim 50$.}
  \label{fgr:bnd}
\end{figure}

\section{Results}
\label{sec3}

\subsection{Structural Properties under Shear}
\label{sec3.1}

For an end-end attraction strength $\varepsilon/k_BT \gtrsim 3$, scaffold structures appear under equilibrium
conditions. \cite{Chelakkot2006,Myung2014,Taslimi2014}
This equilibrium scaffold-like network structure undergoes severe structural rearrangement under shear flow. This is illustrated
in  Fig.~\ref{fgr:shear}, where polymer configurations are shown for the persistence length $L_p/L=1/2$, the end-attraction
strength $\varepsilon/k_BT=4 $, and various shear rates.
As shear flow is applied, the network breaks up and for low shear rates ($Wi \lesssim 5$) densified aggregates are formed. The
scaffold structure persists, but the network phase separates into polymer-rich and polymer-poor domains. At intermediate shear
rates $Wi \approx 2.4$, smaller, partially connected domains are formed, which are reminiscent to micellar structures. \cite{Khalatur1999}
Finally, for high shear rates $Wi \gtrsim 50$, the structural integrity is completely lost and polymers are aligned in a
nematic-like manner along the flow direction.

The initial separation (for $Wi\lesssim 0.5$) into polymer-rich and polymer-poor domains appears in a similar fashion for rodlike polymers.\cite{Taslimi2014} Hence, it seems to be a generic feature of such network structures. However, the shear-induced micellar structures are only observed for more flexible polymers. Here, the flow is sufficiently strong to bend the polymers and induce an attraction between the ends of the same polymer. The nematic alignment at high shear rates is again similar to rodlike polymers. It is caused by the shear forces and appears also for dilute solutions of flexible polymers.\cite{Huang2010,Winkler2010}

To characterize these structures, we determine the average coordination number $\langle
N_{\mathrm{bond}} \rangle$, which is defined as the number of end-beads in proximity of each other, i.e., within distances $r \le 1.5\sigma$.
Figure~\ref{fgr:bond}a shows $\langle
N_{\mathrm{bond}} \rangle$ as a function of the shear rate for the end-attraction strengths $\varepsilon/k_BT = 1.5, \ 3,$ and $4$.  For the lowest value $\varepsilon/k_BT = 1.5$, no  scaffold is formed at equilibrium. \cite{Chelakkot2006,Myung2014,Taslimi2014} In addition, the $\langle N_{\mathrm{bond}} \rangle$ is independent of shear rate, which
indicates that there is no shear-induced network structure either. Naturally, the polymers are aligned by the flow, in a similar fashion as non-attractive polymers. \cite{Huang2010} In systems with scaffold structures, the equilibrium coordination number at zero shear exceeds that of disordered systems considerably, as discussed in more detail in Ref.~\onlinecite{Myung2014}. This equilibrium scaffold structure  is gradually broken by the shear flow for $\varepsilon/k_BT=3 $, and the average coordination number decreases. The number of free ends $N_{\mathrm{end}}$, which are not adjacent to any other end-bead, increases simultaneously. In contrast, for $\varepsilon/k_BT=4 $, the average coordination number first increases with increasing shear rate and passes through a maximum at $Wi \approx 1$. This is associated with the compactification of the scaffold structure visible in Fig.~\ref{fgr:shear}. The attraction is evidently so strong that the shear-induced structural changes lead to an enhanced binding of polymer ends.
The values of $\langle N_{\mathrm{bond}} \rangle$ decrease rapidly with increasing shear rate for $Wi \gtrsim 10$, and the value of an equilibrium non-attractive assembly of polymers is assumed.  Simultaneously, the number of free end-beads $N_{\mathrm{end}}$ increases as the shear rate increases, as shown in Fig.~\ref{fgr:bond}b.

We present the distribution of the coordination number
for various shear rates in Fig.~\ref{fgr:bnd}. The dashed lines are fits to guide the eye, with a Gaussian function for $Wi \lesssim 5$ and an exponential function for $Wi\gtrsim 50$.
Evidently, nodes with a larger number of end-beads are induced at low shear rates ($Wi \lesssim 5$) compare to the distribution without flow ($Wi = 0$). For high shear rates ($Wi \gtrsim 10$), the coordination number is significantly small.
Corresponding mean values are shown in Fig.~\ref{fgr:bond}a.

A qualitative similar behavior of $P(N_{\mathrm{bond}})$ is found for systems at equilibrium and various attraction strengths.\cite{Myung2014} For attraction strengths $\epsilon/k_BT<3$, $P(N_{\mathrm{bond}})$ decreases exponentially with increasing $N_{\mathrm{bond}}$.  For larger values of $\varepsilon$, a maximum of the distribution function appears, as also shown in Fig.~\ref{fgr:bnd}. The exponential decay indicates the  lack of a network structure either due to too weak attraction or too strong external forces.

\begin{figure}
\centering
  \includegraphics[width=8.5cm]{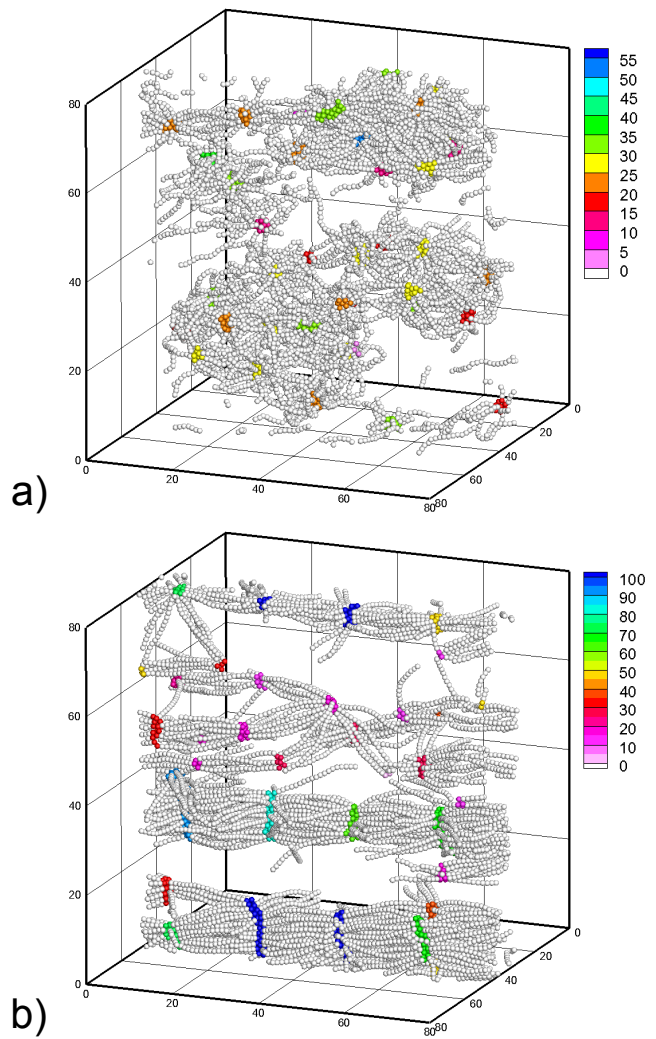}
  \caption{Shear-induced structures at low shear rate ($Wi$) for the
  end-attraction strength $\varepsilon/k_BT=4 $, and different persistence lengths, (a) $L_p/L=1/2$, $Wi=2.4$ and (b) $L_p/L=5$, $Wi=5.1$. Only beads with the slice $30 \le z/a \le 50$ are shown.
  The color code corresponds to the number of adjacent ends. (Multimedia view)}
  \label{fgr:shear_f}
\end{figure}

\begin{figure}
\centering
  \includegraphics[width=8.5cm]{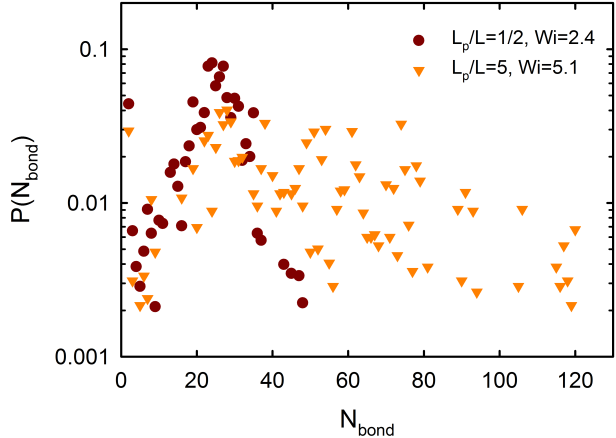}
  \caption{Distributions of the coordination number for the end-attraction strength ${\varepsilon/k_BT}$ = 4.0 and the persistence lengths ${L_p/L=1/2, 5}$.}
  \label{fgr:bnd_f}
\end{figure}

\begin{figure}
\centering
  \includegraphics[width=8.5cm]{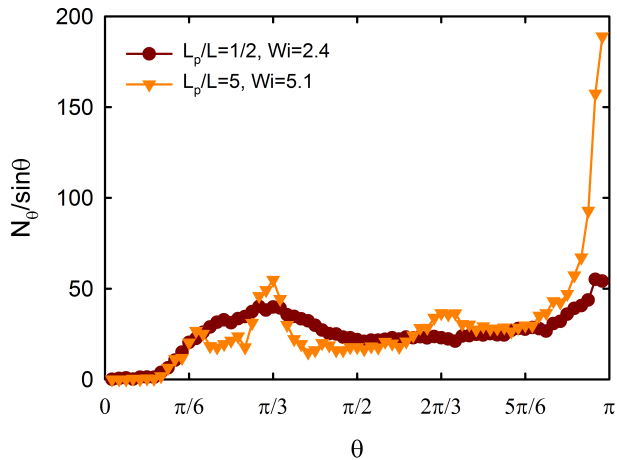}
  \caption{Number of angles ${\theta}$ between bundles for the end-attraction strength ${\varepsilon/k_BT}$ = 4.0 and the persistence lengths ${L_p/L=1/2, 5}$.}
  \label{fgr:theta_f}
\end{figure}

\subsection{Effect of Flexibility}
\label{sec3.2}

Polymer flexibility strongly affects the appearing shear-induced structures. This is reflected in Fig.~\ref{fgr:shear_f} \REV{(Multimedia view)}, where structures are displayed for the persistence lengths $L_p/L=1/2$ and $5$. For semiflexible polymers (${L_p/L=1/2}$), the original scaffold network breaks up and \REV{micellar structures} are formed. In contrast, rodlike polymers (${L_p/L=5}$) are strongly aligned along the flow direction and form thick bundles, an effect already observed for various end-attraction strengths in Ref.~\onlinecite{Taslimi2014}. In both cases,
the end-beads assemble in nodes. For the semiflexible polymers, this can be achieved by significant shear-induced conformational changes of an individual polymer, which gives rise to micellar-like  aggregates. The two ends of a polymer can even meet at the same node. \cite{Khalatur1999,Tabuteau2009,Ligoure2013,Myung2014} This is not possible for rodlike polymers. Their two ends can only participate in two different nodes.\cite{Taslimi2014} In consequence, more dense structures are formed with well aligned rods.
The respective coordination number distributions are shown in Fig.~\ref{fgr:bnd_f}.
Rodlike polymers form nodes with a large number of end-beads, in agreement with the thick bundles (cf. Fig.~\ref{fgr:shear_f}b).

To further characterize the shear-induced structure, Fig.~\ref{fgr:theta_f} presents the distribution $N_{\theta}$ of angles ${\theta}$ between bundles for the two different persistence lengths. Here, a bundle is defined as a connection of two neighboring nodes by two or more polymers.  For semiflexible polymers (${L_p/L=1/2}$), the distribution exhibits a broad peak at ${\theta=\pi/3}$. Note that a peak at ${\theta=\pi/3}$ is a characteristics of a scaffold network, \cite{Chelakkot2006} which is more pronounced in equilibrium structure without flow, \cite{Myung2014} while a peak at ${\theta=\pi}$ indicates parallel alignment of bundles along the flow direction.
For rodlike polymers (${L_p/L=5}$), the peak at ${\theta=\pi}$ is much more pronounced, as expected for bundles. Peaks at ${\theta=\pi/3}$ and ${2\pi/3}$ are also present for rodlike polymers, which implies that the initial scaffold-like connectivity is not completely lost.

\begin{figure}
\centering
  \includegraphics[width=9.0cm]{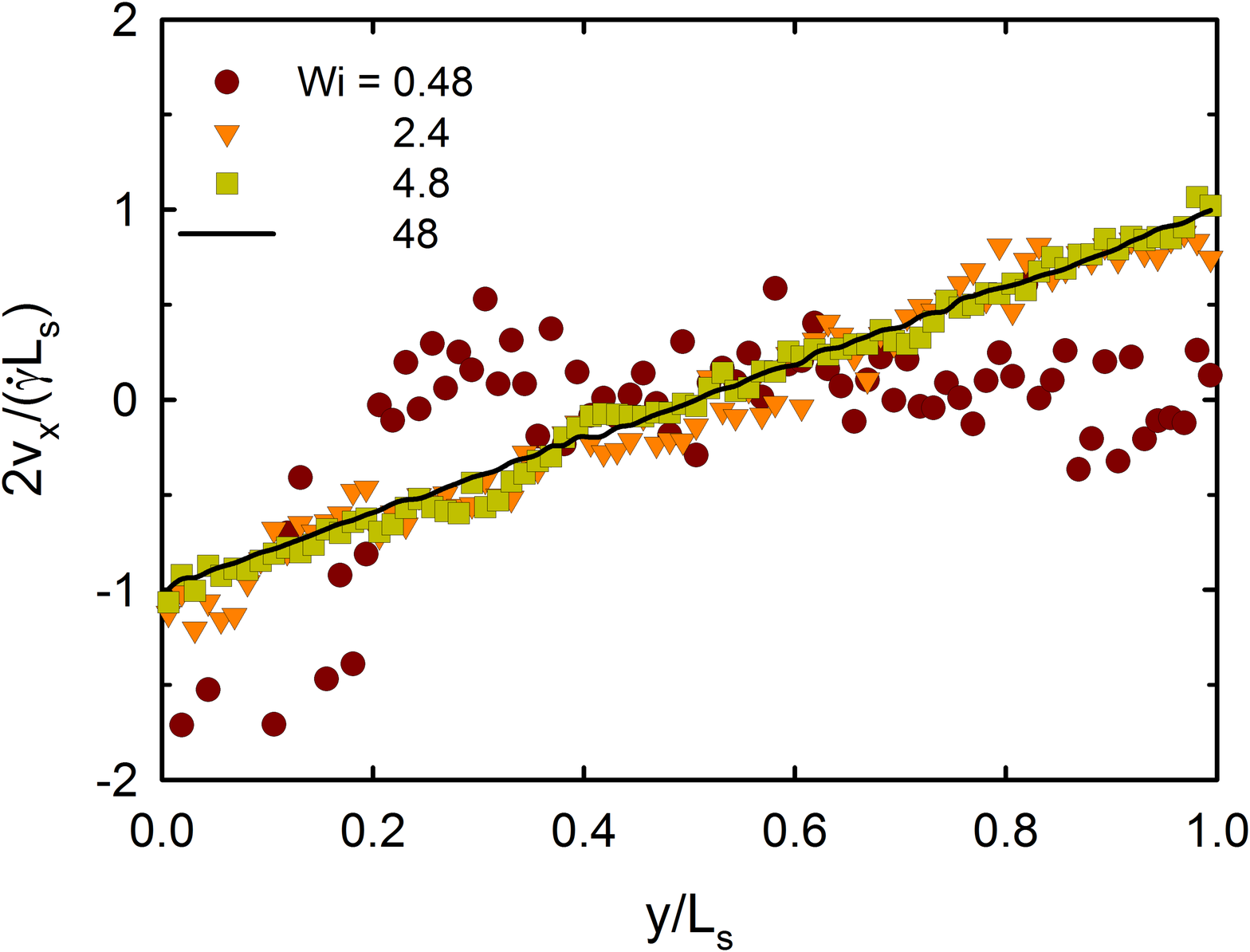}
  \caption{Velocity profiles for the persistence length $L_p/L=1/2$,
  the end-attraction strength $\varepsilon/k_BT=4 $,
  and various shear rates.
  }
  \label{fgr:vel}
\end{figure}

\begin{figure}
\centering
  \includegraphics[width=8.5cm]{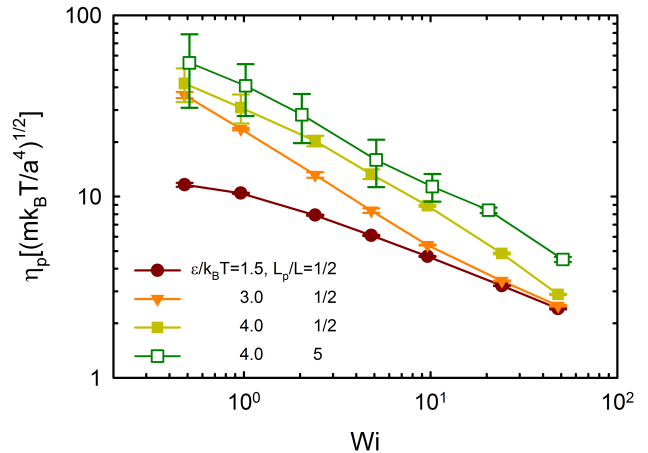}
  \caption{Shear viscosity ($\eta_p$) as a function of shear rate for different persistence lengths and end-attraction strengths.
  }
  \label{fgr:strall}
\end{figure}

\begin{figure}
\centering
  \includegraphics[width=8.5cm]{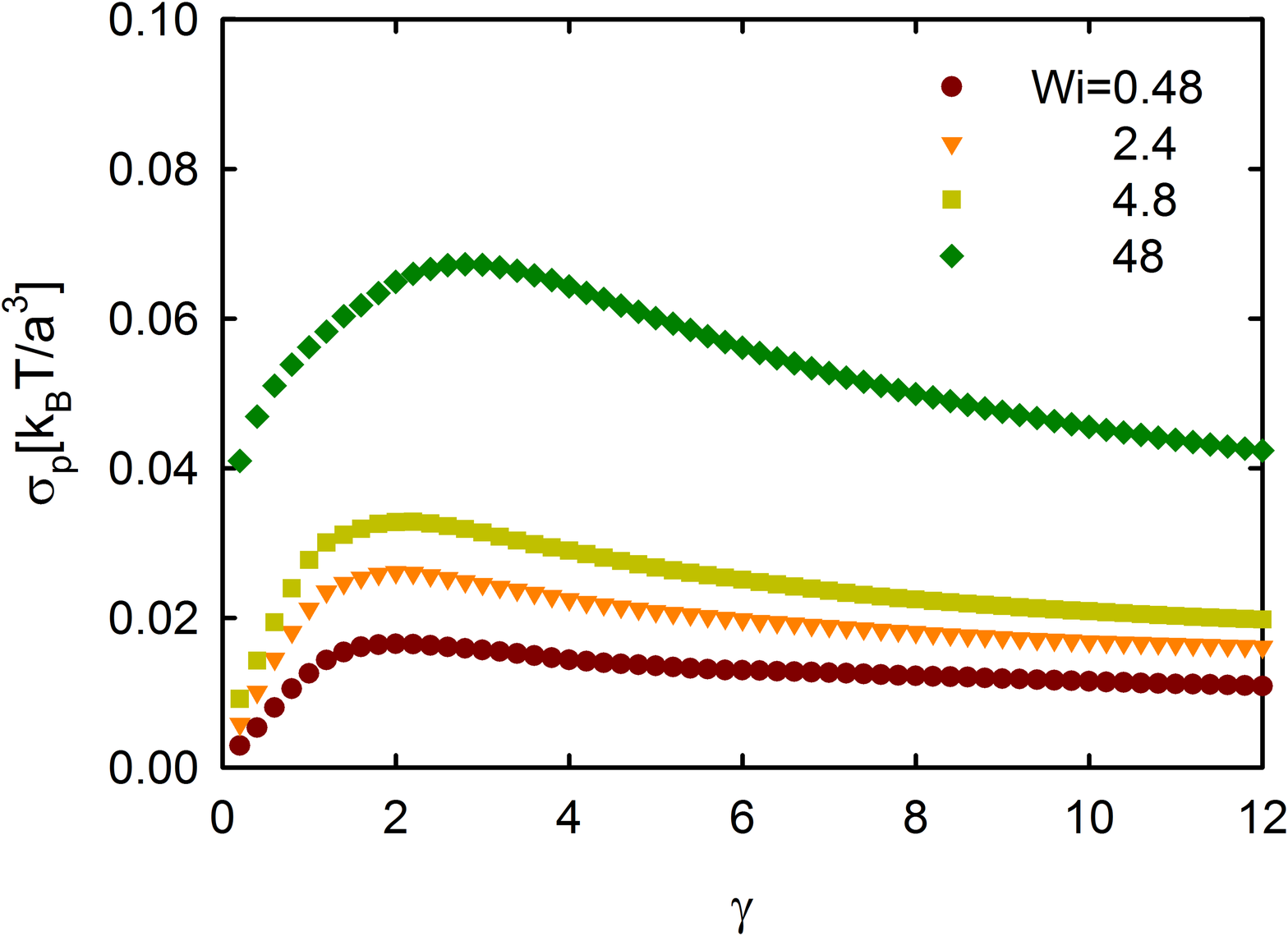}
  \caption{Shear stress as a function of strain ${\gamma = \dot{\gamma}t}$
  in the small-strain region
  for the persistence length $L_p/L=1/2$,
  the end-attraction strength $\varepsilon/k_BT=4 $,
  and various shear rates.
  }
  \label{fgr:str}
\end{figure}

\begin{figure}
\centering
  \includegraphics[width=8.5cm]{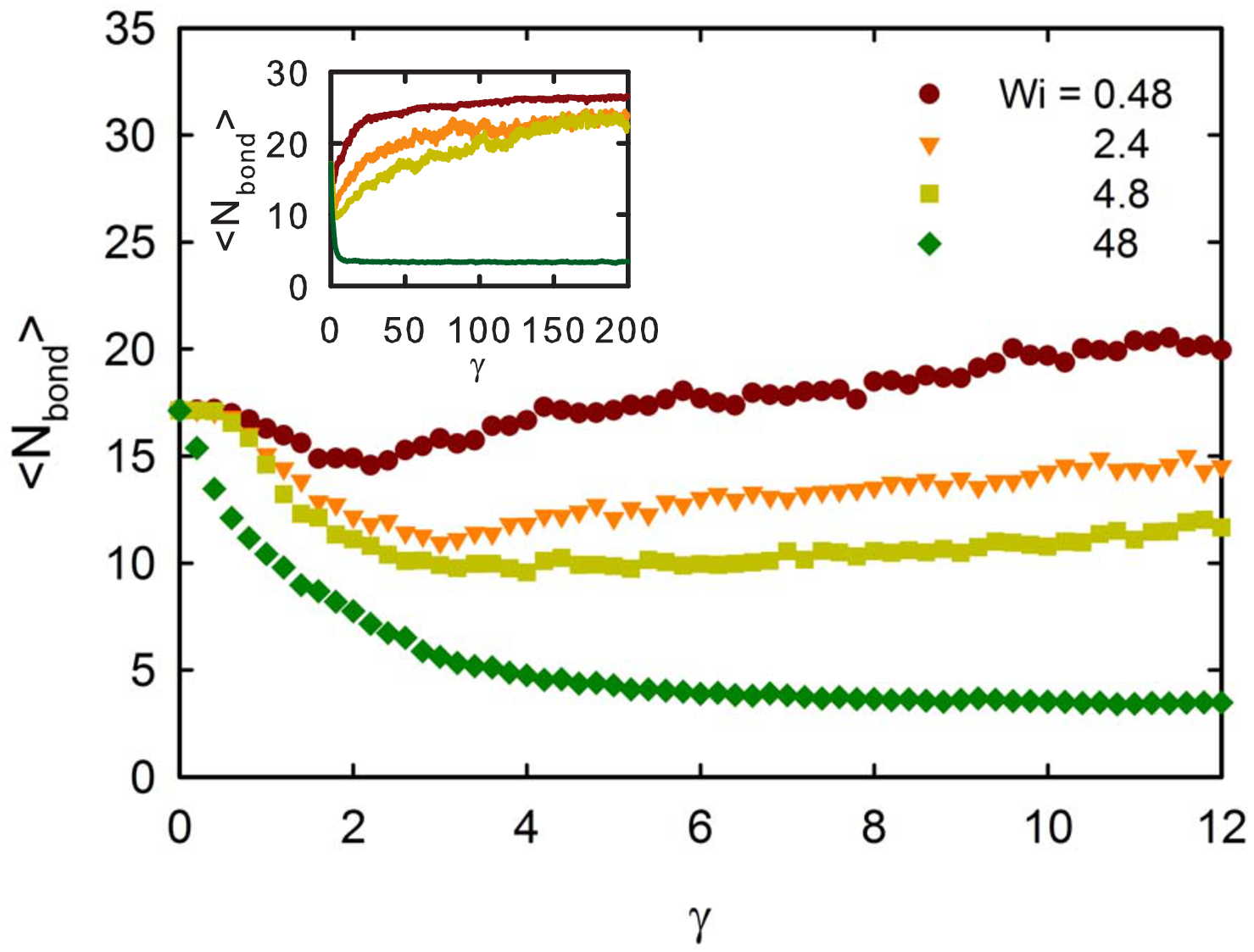}
  \caption{Average coordination number as a function of strain ${\gamma = \dot{\gamma}t}$ in the small-strain region for the persistence length $L_p/L=1/2$,
  the end-attraction strength $\varepsilon/k_BT=4 $,
  and various shear rates. The inset shows the number including large-strain region.
  }
  \label{fgr:bondt}
\end{figure}

\subsection{Rheology} \label{sec3.3}

The structural rearrangement under shear flow affects the rheological properties of the system. \cite{Vermant2005,Sprakel2009}  Figure~\ref{fgr:vel} shows average monomer velocity profiles along the flow-gradient direction for $L_p/L=1/2$ and $\varepsilon/k_BT=4 $.
The flow profiles are non-monotonic for shear rates $Wi \lesssim 5$, which has also been observed in previous studies of rodlike polymers. \cite{Taslimi2014} The bands in the velocity profile can be understood as a consequence of the structural inhomogeneity under shear flow.
The low-shear-rate regions correspond to polymer-rich domains, where a densified network resists the applied shear. In contrast, polymer-poor domains can flow easily, which yields high-shear-rate regions. \REV{For higher shear rates, the velocity profile becomes smoother and we observe a linear monotonic  profile for $Wi =48$. Here,  the structural integrity is lost and polymers are aligned along the flow direction (cf. Fig.~\ref{fgr:shear})}. For both, rodlike\cite{Taslimi2014} and semiflexible polymers,  a monotonic velocity profile is observed for weak end-attraction strengths ($\varepsilon/k_BT<3 $), where the network is either not formed or not strong enough to resist flow.

We present the polymer contribution to the shear viscosity $\eta_p$ as a function of shear rate in Fig.~\ref{fgr:strall}. The polymer contribution to the shear stress $\sigma_p$ is determined by the virial expression
 \begin{align}
 \sigma_p = - \frac{1}{V} \sum_{k} F^p_{x,k} r_{y,k} ,
 \end{align}
where the forces ${\bf F}^p$ follow from the potentials of Eqs.~(\ref{eq:eq3}), (\ref{eq:eq4}), and (\ref{eq:eq5}). \cite{Doi1986,Winkler2009}
The viscosity is then calculated as $\eta_p = \sigma_p/\dot{\gamma}$.
For semiflexible polymers ($L_p/L=1/2$), the viscosity increases with increasing attraction strength for all shear rates (cf. Fig.~\ref{fgr:strall}). In particular, for $\varepsilon/k_BT=4 $, the viscosity of systems of rodlike networks ($L_p/L=5$) is somewhat larger than those comprised of semiflexible polymers ($L_p/L=1/2$). Evidently, the rodlike nature enhances polymer end contacts, and thus, leads to more stable structures. The systems exhibit shear-thinning behavior for the range of applied shear rates, and a Newtonian plateau is observed for weak end-attraction strength ($\varepsilon/k_BT=1.5 $) at low shear rates.

The shear stress $\sigma_p$ in the small-strain region ${\dot{\gamma}t \lesssim 10}$  is plotted in Fig.~\ref{fgr:str}
for $\varepsilon/k_BT=4 $. The stress increases initially in a linear manner. The end of this elastic regime is reached at the strain $\gamma \approx 1$. For larger strains, the network deforms plastically and reaches its maximum strength
for $\gamma \approx 2$. For even larger strains, the stress decreases again. The initial elastic response and yield suggests that there is no  Newtonian viscosity plateau for large attraction strengths.
To shed light on the structural change in
the vicinity of the maximum strength, we present the average coordination number as a function of strain in Fig.~\ref{fgr:bondt}.
Initially ($\gamma < 0.5$), $\langle N_{\mathrm{bond}} \rangle$ is constant for low shear rates ($Wi \lesssim 5$). In this
regime, the network structure is stable and the deformation energy is stored, i.e., the structure behaves elastically.
As strain increases, $\langle N_{\mathrm{bond}} \rangle$ starts to decrease and reaches a minimum at $\gamma \approx 2$,
where the network structure breaks up. For $\gamma > 2$, $\langle N_{\mathrm{bond}} \rangle$ increases again slowly
(cf. inset of Fig.~\ref{fgr:bondt}), which implies that shear-induced aggregates form. For high shear rates ($Wi \gtrsim 10$),
$\langle N_{\mathrm{bond}} \rangle$ decreases monotonically and an asymptotic low steady-state value is assumed.
Here, the network breaks up continuously as shear flow is applied.

\begin{figure}
\centering
  \includegraphics[width=8.5cm]{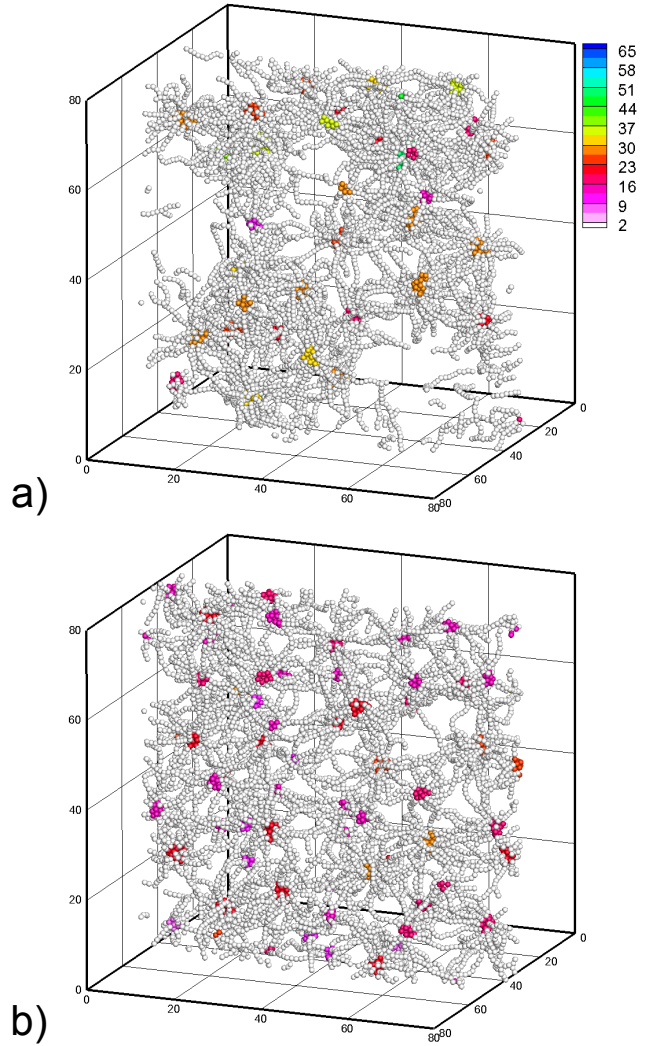}
  \caption{Polymer configurations after cessation of flow for the persistence length $L_p/L=1/2$ and the
  end-attraction strength $\varepsilon/k_BT=4 $. Polymers are relaxed without flow after sheared with (a) $Wi=2.4$ and (b) $Wi=48$.
   Only beads with the slice $30 \le z/a \le 50$ are shown.
  The color code corresponds to the number of adjacent ends.
  }
  \label{fgr:relx}
\end{figure}

\begin{figure}
\centering
  \includegraphics[width=8.5cm]{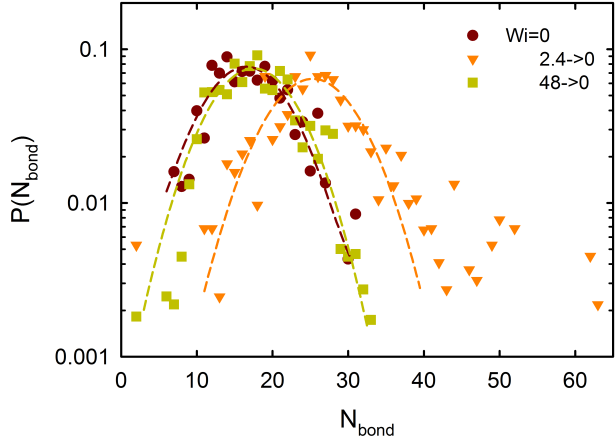}
  \caption{Distributions of the coordination number after cessation of flow for the persistence length $L_p/L=1/2$ and the
  end-attraction strength $\varepsilon/k_BT=4 $. Polymers are relaxed without flow after sheared with different shear rates.
  The dashed lines are fits to guide the eye, with
  a Gaussian function}
  \label{fgr:bnd_relx}
\end{figure}


\subsection{Relaxation after Shear Cessation}
\label{sec3.4}

In order to elucidate the uniqueness of the observed structures, we allow the
shear-induced structures to relax after cessation of flow. Figure~\ref{fgr:relx} shows snapshots of structures after relaxation from initially sheared states for $L_p/L=1/2$ and $\varepsilon/k_BT=4 $. Shear-induced aggregates, which are formed at low shear rates ($Wi = 2.4$) remain after relaxation, and the initial scaffold-like network structure is not fully recovered. When the structural connectivity is fully destroyed for larger shear rates ($Wi = 48$), the system relaxes back to a scaffold-like network. The coordination number distributions for the two structures are shown in Fig.~\ref{fgr:bnd_relx}. In addition, the distribution of $N_{\rm bond}$ of the initial, non-sheared structure is displayed. The recovered structure after high shear rates ($Wi = 48 \to Wi= 0$) shows a similar distribution of the coordination number as the initial scaffold-like network. However, the coordination number is  clearly larger for the relaxed structure after application of a low shear rate ($Wi = 2.4 \to Wi= 0$). Here, a new (equilibrium) structure is formed, which is at least metastable.

Our studies of networks  with weak end-attraction strengths ($\varepsilon/k_BT<3 $) reveal that the  scaffold-like network structure is recovered  after relaxation regardless of pre-applied shear rate. The dependence of the network structure on the initial configuration for strong end-attraction ($\varepsilon/k_BT \ge4 $) has also been observed at equilibrium. \cite{Myung2014} Hence, care has to be taken on equilibrated state of the system.

\section{Conclusion} \label{sec4}

The nonequilibrium structural and dynamical properties of end-functionalized semiflexible polymer suspensions have been investigated by mesoscale hydrodynamic simulations.
Under flow, the scaffold-like network structure of polymers breaks up and densified aggregates are formed at low shear rates, while the structural integrity is completely lost at high shear rates.
We find that network deformation is strongly affected by the polymer flexibility.
Shear-induced aggregates, which are formed at low shear rates and strong end-attraction, show different structures depending on the polymer flexibility.
For semiflexible polymers, the scaffold network breaks up under shear and \REV{micellar structures are formed}. In contrast, rodlike polymers are more strongly aligned along the flow direction and form thick bundles of smectic-like stacks.
For high attraction strengths $\varepsilon/k_BT \gtrsim 4 $, we find that shear-induced dense aggregates remain after relaxation, while the system relaxes back to a scaffold-like network when the structural connectivity is fully destroyed under high shear. For lower attraction strengths, the equilibrium structure is fully recovered.

Our studies shed new light on the nonequilibrium properties of self-organized scaffold structures, specifically their formation and deformation under flow. We expect this knowledge to be useful and provide the basis for further
theoretical and experimental studies of such systems.


%
%

%

\begin{acknowledgments}
Financial support of the project by the EU through
FP7-Infrastructure ESMI (Grant No.262348) is gratefully
acknowledged.
\end{acknowledgments}


%

\end{document}